\documentclass[twocolumn,reprint,amsmath,amssymb, aps,prl]{revtex4-2}
\usepackage[T1]{fontenc}
\usepackage[latin9]{inputenc}
\usepackage{color}
\usepackage{float}

\makeatletter

\usepackage{amsfonts}
\usepackage{latexsym}
\usepackage{epsfig}
\usepackage{graphicx}
\usepackage{tikz}
\definecolor{greatblue}{RGB}{40,120,181}
\definecolor{greatred}{RGB}{200,36,35}
\usepackage[colorlinks,linkcolor=greatblue,anchorcolor=blue,citecolor=greatred]{hyperref}

\usepackage{dcolumn}
\usepackage{bm}

\makeatother

\usepackage[english]{babel}

\begin{document}
\preprint{preprintnumbers}{CTP-SCU/2024013}

\title{How Einstein's Equation Emerges From CFT$_2$}

\author{Xin Jiang}
\email{domoki@stu.scu.edu.cn}
\affiliation{College of Physics, Sichuan University, Chengdu, 610065, China}

\author{Peng Wang}
\email{pengw@scu.edu.cn}
\affiliation{College of Physics, Sichuan University, Chengdu, 610065, China}

\author{Houwen Wu}
\email{iverwu@scu.edu.cn}
\affiliation{College of Physics, Sichuan University, Chengdu, 610065, China}

\author{Haitang Yang}
\email{hyanga@scu.edu.cn}
\affiliation{College of Physics, Sichuan University, Chengdu, 610065, China}

\date{\today}

\begin{abstract}
The {\it finiteness}  of the  entanglement entropies  between disjoint subsystems
enables us to show that,  the dynamical equation of
the entanglement entropy in CFT$_2$  is precisely three dimensional Einstein's equation.
We establish a profound relation between the cosmological constant and CFT$_2$  entanglement entropy.
Thus   entanglement entropies  induce  internal  gravitational geometries in CFT$_2$.
Extracting the dual metric from an entanglement entropy becomes a straightforward procedure.
Remarkably, we discover that the renormalization group equation  is a geometric identity.

\end{abstract}

\maketitle

\section*{Introduction}

The AdS/CFT correspondence  \cite{Maldacena:1997re} has dominated high energy theory over the past three decades.
As a testable realization for the holographic principle \cite{tHooft:1993dmi}, it conjectures that a
weakly coupled gravitational theory in the bulk of AdS$_{d+1}$ is equivalent to a strongly coupled
CFT$_d$ on the conformally flat boundary. Among numerous developments, a central subject is
to reconstruct  gravity from the dual QFT, or ``derive'' Einstein's equation in  QFT.

Since Ryu and Takayanagi (RT) identified the entanglement entropy between adjacent regions  of  CFT$_d$ 
with the area of  bulk  minimal surfaces anchored on the  conformal boundary of AdS$_{d+1}$ \cite{Ryu:2006bv,Ryu:2006ef},
among many approaches, using entanglement entropy  to generate the dual geometry has become
a very promising direction. For $d=2$ considered in this letter,  the minimal surfaces are geodesics.

In   an  entangled  pure CFT state $\psi_{A\bar A}$, entanglement entropy classifies quantum entanglement between complementary parts $A$ and $\bar A$. It is defined by the von Neumann entropy: $S_{\text{vN}}(A)= -\mathrm{Tr} \rho_{A}\log \rho_{A}$. Here $\rho_{A}$ is  a reduced density matrix $\rho_{A}=\mathrm{Tr}_{\bar A}\vert\psi_{A\bar A}\rangle\langle\psi_{A\bar A}\vert$. 
However, as the entangled complementary subsystems are continuously connected, 
the  entanglement entropy diverges due to the very intense entanglement
between neighboring fields. This asymptotic  behavior makes it very difficult, 
if not completely impossible, to extract certain exact relations.

It is crucial to note that, the   entanglement entropy defined above  fails to  characterize the  entanglement
in a mixed state $\rho_{AB}$.
The traditional resolution is to purify the mixed state  by adding  auxiliary
systems $\bar{A}\bar{B}$. Then the  entanglement of purification (EoP or $E_{P}$) \cite{Terhal:2002riz}
is defined by minimizing  $S_{\text{vN}}(A\bar{A}:B\bar B)$ over all possible
purifications,  and quantifies the entanglement between $A$ and $B$:
$S_{\mathrm{vN}}(A:B) :=E_P = \min_{\bar A\bar B} S_{\text{vN}}(A\bar{A}:B\bar B)$ \cite{Takayanagi:2017knl}.
However, optimization over purifications is, in practice, intractable in CFTs.

Recently, the SUBTRACTION approach is proposed to calculate the entanglement entropy $S_{\mathrm{vN}}\left(A:B\right)$
between disjoint subsystems $A$ and $B$ in CFT$_2$ \cite{Jiang:2024ijx,Jiang:2025tqu,Jiang:2025dir}.
This quantity can be equivalently interpreted as the entanglement entropy in the mixed state $\rho_{AB}$, as explained in the appendix.
The disconnectedness between subsystems $A$ and $B$ makes $S_{\mathrm{vN}}\left(A:B\right)$ finite.
This enables us to explicitly  show how three dimensional Einstein's gravity emerges from CFT$_2$ in this letter.
Then we will demonstrate that the renormalization group (RG) equation is surprisingly a geometric identity.

\section*{The entanglement entropy for  disjoint subsystems in CFT$_2$}
\begin{figure}[h]
\includegraphics[width=0.3\textwidth]{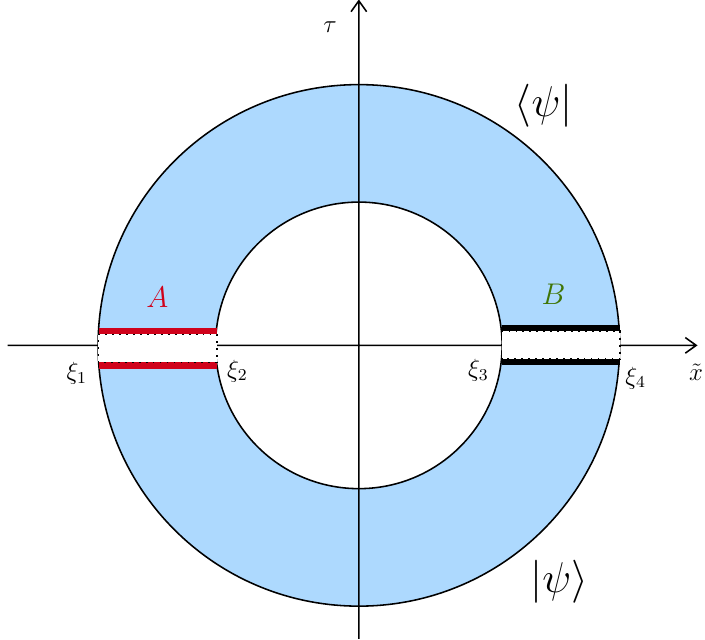}
\caption{The Euclidean pure state density matrix $\rho=|\psi\rangle\langle\psi|$ for the annular  CFT$_2$. $\xi_i = \tilde x_i + i\tilde t_i$ with $\tilde t_i = - i\tau_i$.
\label{fig:density}}
\end{figure}

To start with, consider the annular CFT$_2$. We want to calculate the entanglement entropy between
distinct intervals. The pure state density matrix is shown in Figure \ref{fig:density}.

The entanglement entropy $S_{\text{vN}}(A)$ between $A$ and $B$ is given by
\begin{equation}
S_{\text{vN}}(A)  =  \lim_{n\rightarrow1} \frac{1}{1-n}\log\mathrm{Tr}_{A}\rho_{A}^{n},
\label{eq:Renyi}
\end{equation}
where  $\rho_A= \mathrm{Tr}_{B} \rho$ and $\mathrm{Tr}_{A}\rho_{A}^{n}=\frac{Z_{n}}{Z_{1}^{n}}$.
Here, $Z_{1}$ is the partition function on the
annulus. $Z_{n}$ is the partition function on the $n$-sheeted
cover $\mathcal{M}_{n}$ obtained by cyclically gluing  $n$ copies of annuli along $A$.
$S_{\text{vN}}(A)$  for a static time slice is given in \cite{Jiang:2024ijx}.
For generic time dependent asymmetric cases,
$\xi_i = \tilde x_i + i\tilde t_i$ and $\tilde t_i = -i \tau_i$ for Euclidean time,  $S_{\text{vN}}(A)$ has been
calculated in \cite{Jiang:2025tqu},
\begin{equation}
S_{\text{vN}}(A)  =  \frac{c}{12}\log\left(\frac{1+\sqrt{\eta}}{1-\sqrt{\eta}}\right)+\frac{c}{12}\log\left(\frac{1+\sqrt{\bar{\eta}}}{1-\sqrt{\bar{\eta}}}\right),
\label{eq:Asymmetric-EE}
\end{equation}
with the cross ratio
\begin{equation}
\eta=\frac{\xi_{21}\xi_{43}}{\xi_{31}\xi_{42}}, \quad \xi_{ij}\equiv \xi_i -\xi_j.
\label{eq:cross-ratio}
\end{equation}

As a consequence of the RG equation $\ell \frac{d}{d\ell} S = \frac{c}{6}$ and conformal symmetries,
the entanglement entropy (\ref{eq:Asymmetric-EE}) turns out to be the universal expression for CFT$_2$s.
Different CFTs, such as a finite temperature system or a finite size system,
are characterized by different cross ratios \cite{Jiang:2025tqu}.

\section*{Einstein's equation from entanglement entropies in CFT$_2$}

The endpoints uniquely fix $A$
and $B$, which amounts to eight free real parameters. However, two of these are gauge freedoms as shown in \cite{Jiang:2025tqu}. So, it suffices to consider $A$ and $B$ being collinear, in which six real numbers fix $A$ and $B$.
The disjoint entanglement entropy (\ref{eq:Asymmetric-EE}) can be expressed in terms of any equivalent set of
linear independent parameters. The physics behind is the same. For convenience, we regroup the   parameters as,
\begin{eqnarray}
x+it &=&\xi_{1}+\frac{\vert\xi_{13}\vert\vert\xi_{14}\vert}{\vert\xi_{12}\vert\vert\xi_{13}\vert+\vert\xi_{34}\vert\vert\xi_{24}\vert}\xi_{21},\nonumber\\
x'+it'&=&\xi_{2}+\frac{\vert\xi_{23}\vert\vert\xi_{24}\vert}{\vert\xi_{13}\vert\vert\xi_{34}\vert+\vert\xi_{12}\vert\vert\xi_{24}\vert}\xi_{21},\nonumber\\
z&=&\frac{\vert\xi_{14}\vert\sqrt{\vert\xi_{12}\vert\vert\xi_{13}\vert\vert\xi_{34}\vert\vert\xi_{24}\vert}}{\vert\xi_{12}\vert\vert\xi_{13}\vert+\vert\xi_{34}\vert\vert\xi_{24}\vert},\nonumber\\
z'&=&\frac{\vert\xi_{23}\vert\sqrt{\vert\xi_{12}\vert\vert\xi_{13}\vert\vert\xi_{34}\vert\vert\xi_{24}\vert}}{\vert\xi_{13}\vert\vert\xi_{34}\vert+\vert\xi_{12}\vert\vert\xi_{24}\vert},
\label{eq:parameter-redefinitions}
\end{eqnarray}
The cross ratio becomes a real number,
\begin{equation}
\eta=\left|\frac{\xi_{21}\xi_{43}}{\xi_{31}\xi_{42}}\right|
 = \frac{\Delta z^{2}+\Delta x^2 -\Delta t^2}{(z+z')^{2}+ \Delta x^2 -\Delta t^2}.
	\label{eq:cross-ratio-x}
\end{equation}
In terms of these new coordinates $(t,x,z)$ and $(t',x',z')$, the entanglement entropy $S_{\text{vN}}(A)$ can be written as
\begin{equation}
S_{\text{vN}}(A)=\frac{c}{6}\cosh^{-1}\left(1+\frac{\Delta z^{2}+\Delta x^2 -\Delta t^2}{2zz'}\right).
\label{eq:EE-x}
\end{equation}
We define a metric
\begin{equation}
\chi(x,x'):=\frac{1}{2} S_{\mathrm{vN}}^2 (A),\,\,\, g_{\mu\nu}=-[\chi_{\mu\nu'}]:  =- \lim_{x'\to x} \partial_{\nu'}\partial_\mu \chi,
\label{eq:chi}
\end{equation}
where the bracket $[\cdots]$ denotes the coincidence limit $x\to x'$. Indices in  brackets are derivatives with respect to $x^\mu$ or ${x'}^\mu$ hereinafter. Introduce
\begin{equation}
R_{\mu\nu}:=g^{\rho\sigma} [\nabla_\nu\nabla_\sigma\nabla_\mu\partial_\rho \chi],\quad R:= g^{\mu\nu} R_{\mu\nu},
\label{eq:definitions}
\end{equation}
where the covariant derivatives $\nabla_{(\centerdot)}$ are defined by $g_{\mu\nu}$.
Then, plugging in everything, it is straightforward to verify
\begin{equation}
R_{\mu\nu}-\frac{1}{2}g_{\mu\nu} R + \Lambda_c g_{\mu\nu}=0.
\label{eq:EE-EOM}
\end{equation}
This is precisely the three dimensional Einstein's equation with cosmological constant,
\begin{equation}
\Lambda_c = \frac{R}{6} =\frac{1}{6} \Big[\chi^{\mu\nu'} \chi^{\rho\sigma'} \chi_{\rho\mu\sigma\nu} \Big].
\label{eq:Lambda}
\end{equation}

\noindent So, the cosmological constant is not really a free parameter, but determined by the CFT entanglement entropy.
From equation (\ref{eq:chi}), extracting the metric from the entanglement entropy is straightforward, whereas
this had been a highly non-trivial task when using the divergent entanglement entropies of connected regions \cite{Wang:2017bym,Wang:2018vbw,Wang:2018jva}.

As we explained at the end of the last section, CFT$_2$s on other topologies possess the same entanglement entropy expression (\ref{eq:Asymmetric-EE}) with different cross ratios which are related by conformal transformations.  Since equations (\ref{eq:chi}-\ref{eq:Lambda})
are all tensorial, we immediately conclude that the parameter regrouping in equation (\ref{eq:parameter-redefinitions})
is \emph{once-for-all} applicable to all CFT$_2$s and the  derivation applies to all  CFT$_2$s.
In a companion work \cite{Jiang:2024xqz}, we use the thermal CFT as an explicit example to realize ER=EPR.

We therefore derived Einstein's equation from the quantum entanglement of CFT$_2$!

\section*{Intuitive understanding of the derivation}

Although our derivation of Einstein's equation is entirely within the framework of CFT, 
it is  instructive to provide an intuitive picture.
To this end, let us turn to geometry. Consider a   geodesic $x(\tau), \tau\in [0,t]$ connecting two points $x(\tau=0)$ and $x'(\tau=t)$.
$L(x,x')$ is the geodesic length. The Synge's world function  is a biscalar of $x$ and $x'$,
defined as,
\begin{equation}
\sigma(x,x'): = \frac{1}{2} L^2(x,x') =  \frac{1}{2} \Delta\tau
\int_0^t d\tau g_{\mu\nu} \frac{dx^\mu}{d\tau} \frac{dx^\nu}{d\tau}.
\label{eq:SyngeFunction}
\end{equation}
The metric turns out to be \cite{Synge:1960ueh}
\begin{equation}
g_{\mu\nu}= -[\sigma_{\mu\nu'}]:= -\lim_{x'\to x} \partial_{\nu'} \partial_{\mu} \sigma (x,x').
\label{eq:GtoM1}
\end{equation}
By default, the indices of $\sigma$ are ordinary or covariant  derivatives.
With respect to  metric, we have Einstein's  equation
\begin{equation}
R_{\mu\nu} -\frac{1}{2}g_{\mu\nu}\,R  +\Lambda g_{\mu\nu} =0.
\label{eq:EinteinE}
\end{equation}
We only address vacuum with  $\Lambda\not=0$. Since
\begin{equation}  [\sigma_{\alpha\mu\beta\nu}]:= \lim_{x'\to x} \nabla_\nu\nabla_\beta \nabla_\mu\partial_\alpha \sigma = \frac{ R_{\alpha\nu\beta\mu}+ R_{\alpha\beta\nu\mu}}{3},
\label{eq:SigmaR}
\end{equation}
one finds
\begin{equation}
\Sigma_{\mu\nu}:= [\sigma^{\alpha\beta}\, \sigma_{\alpha\mu\beta\nu}]  =g^{\alpha\beta}  [\sigma_{\alpha\mu\beta\nu}]  = \frac{1}{3} R_{\mu\nu},
\end{equation}
and
\begin{equation}
\Sigma:= [\sigma^{\mu\nu} \sigma^{\alpha\beta}\, \sigma_{\alpha\mu\beta\nu}]= g^{\mu\nu}\Sigma_{\mu\nu} = \frac{1}{3} R.
\end{equation}
Note raising or lowering indices and taking the coincidence limit commute. So, the geodesic version of Einstein's equation is
\begin{equation}
\Sigma_{\mu\nu}-\frac{1}{2}[\sigma_{\mu\nu}]\,\Sigma  +\frac{\Lambda}{3} [\sigma_{\mu\nu}]=0.
\label{eq: EOM-of-Geodesics}
\end{equation}

\begin{figure}[h]
\includegraphics[width=0.5\textwidth]{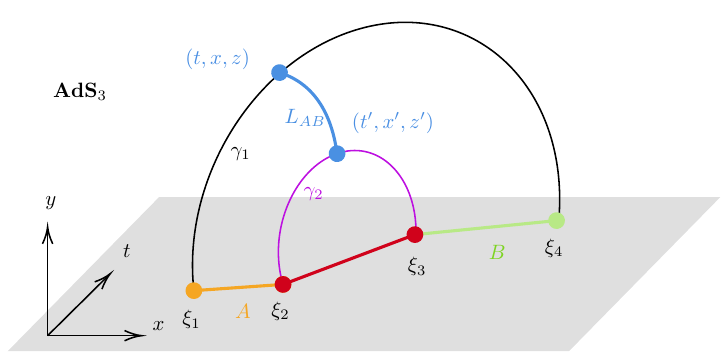}
\caption{In AdS$_3$, geodesic $\gamma_1$ is completely fixed by $\xi_1$ and $\xi_4$, geodesic $\gamma_2$ is completely fixed by $\xi_2$ and $\xi_3$. $L_{AB}$ is a geodesic and unique.
\label{fig:geodesic}}
\end{figure}

Now considering AdS$_3$ geometry, on the conformal boundary, two segments $A$ and $B$ are disjoint.
Referring to Figure \ref{fig:geodesic}, $\gamma_1$ and $\gamma_2$  are both geodesics. From the Ultra-parallel theorem \cite{Buser2010} (Theorem 1.1.6) in hyperbolic geometry,
the shortest curve connecting $\gamma_1$ and $\gamma_2$ is a unique geodesic $L_{AB}$. So, $(t,x,z)$ and $(t',x',z')$ are completely fixed by the four $\xi_i$. The picture also demonstrates explicitly how an extra dimension arises from the boundary theory: four points on the boundary lead to two points in the bulk.
Note four points on the boundary have eight parameters.
However, two of these turn out to be gauge parameters  \cite{Jiang:2025tqu}. The remaining six independent
parameters exactly match  two points in the 3-dimensional bulk.

Since $L_{AB}$ has finite length, it must satisfy  Einstein's equation (\ref{eq: EOM-of-Geodesics}). On the other hand,
one can easily calculate that the length of $L_{AB}$  is precisely equal to $S_{\text{vN}}(A)$ in equation (\ref{eq:EE-x}) with the Brown-Henneaux's formula \cite{Brown1986} $c=3R_{\text{AdS}}/2G^{(3)}$ assumed. This is nothing but the renowned RT formula \cite{Ryu:2006bv,Ryu:2006ef}. 
Therefore, we reached a holographic description of the CFT Einstein's equation. We further identified $\Lambda_c = \Lambda/3$. Note this holographic description is only for illustrative purposes. It is not a necessary ingredient in our derivation.

A key point to emphasize is that this work is entirely guided by the RT formula. From our derivation, only a finite entanglement entropy is required.
There exist several finite measures of entanglement, including the odd entanglement entropy \citep{Tamaoka:2018ned}, reflected entropy \citep{Dutta:2019gen}, balanced partial entanglement \citep{Wen:2021qgx}, and distillable entanglement \citep{Mori:2024gwe}. Theoretically, any of these could be used to derive the Einstein equation. However, taking reflected entropy \citep{Dutta:2019gen} as an example, its holographic dual is a closed curve formed by gluing two identical geodesics. This structure makes it impossible to perform endpoint derivatives and take the coincidence limit. 
So, unless 
obtaining the exact CFT$_2$ dual, i.e. $S_{\text{vN}}(A)$ in equation (\ref{eq:Renyi})
of the geodesic $L_{AB}$ in Figure \ref{fig:geodesic}, we could not realize how to derive Einstein equation from the entanglement entropy.

\section*{RG equation and Klein-Gordon equation}

On the QFT side, we have an entanglement generated gravity in addition to the CFT itself.  Whereas on the gravity side, we have spacetime geometry and QFT living on it. The spacetime geometry and entanglement generated gravity have identical structures.

Basically, there are three fundamental equations, Einstein's equation, covariant Klein-Gordon equation and RG equation.
We already know that in bulk gravity, Einstein's equation and covariant Klein-Gordon equation exist, whereas in the CFT   on the boundary, we have quantum entanglement generated Einstein's equation and RG equation. Therefore, we want to find the  covariant Klein-Gordon equation in CFT and RG equation in gravity.

\vspace{2ex}
\noindent\underline{RG equation as a geometric identity}
\vspace{2ex}

For a classically scale-invariant theory living on a $d+1$-dimensional
manifold $\mathcal{M}$ with the metric $ds^{2}=\gamma_{ab}dx^{a}dx^{b}$,
the Callan-Symanzik RG equation is \cite{Osborn:1991gm}:
\begin{equation}
\left[\ell\frac{\partial}{\partial\ell}-2\int_{\mathcal{M}}\gamma^{ab}\frac{\delta}{\delta\gamma^{ab}}\right]\log Z_{\text{CFT}}=0,
\end{equation}
where $\ell$ is the length scale.
Using the replica trick \cite{Calabrese:2004eu} and R\'{e}nyi entropy
\begin{equation}
S_{\text{vN}}^{(n)}=\frac{1}{1-n}\log\left[\frac{Z_{\mathcal{M}_{n}}}{\left(Z_{\mathcal{M}}\right)^{n}}\right],
\end{equation}
where $\mathcal{M}_{n}$ is the replicated manifold,
one can show the RG equation for the entanglement entropy is \cite{Jiang:2023loq}
\begin{equation}
\ell\frac{\partial}{\partial\ell}S_{\text{vN}}=\frac{c}{6}.
\label{eq:S_RG}
\end{equation}
The constant $\frac{c}{6}$ on the right hand side is  for $d=2$.

On the other hand, in {\it any} geometry, not limited to gravity, there is an identity for the world function (\ref{eq:SyngeFunction}),
\begin{equation}
g^{\mu\nu}\sigma_\mu \sigma_\nu = 2\sigma, \quad  \sigma(x,x') = \frac{1}{2} L^2(x,x').
\label{eq:identity}
\end{equation}

On a time slice of AdS$_3$,  the isometry is  SL$(2,R)$.
Those isometries preserve the distance between points.
Except the vertical ones, any other geodesic is a segment of a  semi-circle,
\begin{equation}
(x-x_{0})^{2}+z^{2}=r^{2}.
\end{equation}
We can map a semi-circle to a  vertical line by the following isometries
\begin{equation*}
(\tilde x, \tilde z):= (\frac{x-x_{0}+r}{(x-x_{0}+r)^{2}+z^{2}}-\frac{1}{2r}, \frac{z}{(x-x_{0}+r)^{2}+z^{2}}).
\end{equation*}
In terms of $(\tilde{x},\tilde{z})$, the  semi-circle  is a vertical line ${\sigma}={\sigma}(0,\tilde{z}_{1};0,\tilde{z}_{2})$.
Now, apply the geometric  identity (\ref{eq:identity}),
\begin{equation}
z^{2}\left[(\partial_{x}\sigma)^{2}+(\partial_{z}\sigma)^{2}\right]=\tilde{z}^{2}(\partial_{\tilde{z}}{\sigma})^{2}=2{\sigma}.
\end{equation}
Obviously, $\tilde{z}$ is the renormalization scale $\ell$.
Including the coupling $c=3R_{\text{AdS}}/2G^{(3)}$, in either the quantum entanglement generated gravity in CFT$_2$ or the bulk spacetime gravity,  $\sigma = \frac{1}{2} L^2 =\frac{1}{2} S_{\text{vN}}^2$, we thus have
in both  boundary CFT$_2$ and bulk geometry,
\begin{equation}
\ell \frac{\partial}{\partial\ell} S_{\text{vN}} = \frac{c}{6},
\end{equation}
precisely the RG equation (\ref{eq:S_RG}).

\vspace{2ex}
\noindent\underline{Covariant Klein-Gordon equation in CFT}
\vspace{2ex}

Usually  the entanglement entropy is computed by using the replica method. The  R\'{e}nyi entropy is determined by the partition function of the replicated manifold. The replicated partition function equals the correlation function of twisted operators,

\begin{equation}
G\left(u(\xi_i),v(\xi_i)\right)=\mathrm{Tr}_{A}\rho_{A}^{n}.
\end{equation}
In \cite{Jiang:2024ijx}, we have obtained $\mathrm{Tr}_{A}\rho_{A}^{n}=e^{\frac{1}{2}(\frac{1}{n}-n) S_{\mathrm{vN}}\left(A:B\right)}$. So,
\begin{equation}
G\left(u(\xi_i),v(\xi_i)\right)=e^{\frac{1}{2}(\frac{1}{n}-n) S_{\mathrm{vN}}\left(A:B\right)}.
\label{eq:correlation}
\end{equation}
Note $u\left(\xi_i\right)$ and $v\left(\xi_i\right)$ are basically $(t,x,z)$ and $(t',x',z')$ respectively in Fig. (\ref{fig:geodesic}), but located on the boundary of the cut geometry.  Thus $G\left(u(\xi_i),v(\xi_i)\right)$ is not
the usual CFT propagator but effectively a four point function.

To obtain the EOM of this propagator, we refer to  the AdS$_3$ bulk propagator of a scalar with mass
$m^2=\triangle\left(\triangle-2\right)$ and  $\triangle =\frac{c}{12}(n-1/n)$,
\begin{equation}
G_{\rm Bulk}\left(u,v\right)
=\frac{e^{-\triangle L\left(u,v\right)}}{1-e^{-2L\left(u,v\right)}},
\end{equation}
which satisfies the bulk covariant Klein-Gordon equation.
Under the semiclassical limit, $\triangle\rightarrow\infty$, keeping $\triangle/c$ fixed,
this propagator reduces to the CFT one (\ref{eq:correlation}). Therefore, we have
\begin{equation}
(\nabla^2 - m^2) G\left(u(\xi_i),v(\xi_i)\right) = \delta^3_{\cal M}(\xi_i-\xi_j),
\end{equation}
at large $c$ limit.

\section*{Summary and discussion}
We demonstrated
that the dynamics of  entanglement entropy in CFT$_2$ are precisely three dimensional  Einstein's  equation.
A profound relation between the cosmological constant and CFT entropy is also given.
We thus uncovered a hidden gravitational structure within CFT$_2$. We confirmed that in addition to Einstein's equation and covariant Klein-Gordon equation, in the bulk, a geometrical identity is the RG equation. On the CFT side, in addition to the quantum entanglement generated Einstein's  equation,  we gave the  covariant Klein-Gordon equation,
and showed that the RG equation is a geometric identity of the emerged  gravity.

Since the ordinary QFT and the induced gravity are linked through entanglement entropy,
it is now possible to unify the black hole thermodynamics and Einstein's equation.

In this paper, we have focused on AdS. Nevertheless, the derivation is applicable to dS with the same fashion.

We want to emphasize some 
important features indicated by the formula   $g_{\mu\nu} = -[\frac{1}{2}(S^2_{\rm vN})_{\mu\nu'}]$:
\begin{itemize}
\item We showed that the metric is a derived but not a fundamental quantity. Perhaps this is  why gravity is non-renormalizable and we should not try to quantize the metric itself. Furthermore, $S_{\rm vN}$ is already a quantum quantity, why bother perform further
quantization?
\item Note the explicit local/global duality as shown by the formula.
\item Obviously spacetime singularities are phase transitions from this formula.
\item The LHS is geometric, whereas the RHS is algebraic.
\end{itemize}
We addressed the vacuum Einstein's equation in CFT$_2$. Including  matter sources is absolutely a critical  question. 
Perhaps,  the most important future direction is to study CFT$_d$ for $d>2$. 


In physics, energy dominates and links everything. But now, we tend to believe the fundamental object in physics is entropy.

\vspace*{3.0ex}
\begin{acknowledgments}
\paragraph*{Acknowledgments.}
This work is supported in part by NSFC (Grant No. 12105191, 12275183 and 12275184).
\end{acknowledgments}

\bibliographystyle{unsrturl}
\bibliography{ref202504}

\begin{thebibliography}{10}

\bibitem{Maldacena:1997re}
Juan~Martin Maldacena.
\newblock {The Large N limit of superconformal field theories and
  supergravity}.
\newblock {\em Adv. Theor. Math. Phys.}, 2:231--252, 1998.
\newblock \href {http://arxiv.org/abs/hep-th/9711200}
  {\path{arXiv:hep-th/9711200}}, \href
  {http://dx.doi.org/10.1023/A:1026654312961}
  {\path{doi:10.1023/A:1026654312961}}.

\bibitem{tHooft:1993dmi}
Gerard 't~Hooft.
\newblock {Dimensional reduction in quantum gravity}.
\newblock {\em Conf. Proc. C}, 930308:284--296, 1993.
\newblock \href {http://arxiv.org/abs/gr-qc/9310026}
  {\path{arXiv:gr-qc/9310026}}.

\bibitem{Ryu:2006bv}
Shinsei Ryu and Tadashi Takayanagi.
\newblock {Holographic derivation of entanglement entropy from AdS/CFT}.
\newblock {\em Phys. Rev. Lett.}, 96:181602, 2006.
\newblock \href {http://arxiv.org/abs/hep-th/0603001}
  {\path{arXiv:hep-th/0603001}}, \href
  {http://dx.doi.org/10.1103/PhysRevLett.96.181602}
  {\path{doi:10.1103/PhysRevLett.96.181602}}.

\bibitem{Ryu:2006ef}
Shinsei Ryu and Tadashi Takayanagi.
\newblock Aspects of holographic entanglement entropy.
\newblock {\em Journal of High Energy Physics}, 2006(08):045--045, aug 2006.
\newblock \href {http://dx.doi.org/10.1088/1126-6708/2006/08/045}
  {\path{doi:10.1088/1126-6708/2006/08/045}}.

\bibitem{Terhal:2002riz}
Barbara~M. Terhal, Michal Horodecki, Debbie~W. Leung, and David~P. DiVincenzo.
\newblock {The entanglement of purification}.
\newblock {\em J. Math. Phys.}, 43(9):4286--4298, 2002.
\newblock \href {http://arxiv.org/abs/quant-ph/0202044}
  {\path{arXiv:quant-ph/0202044}}, \href {http://dx.doi.org/10.1063/1.1498001}
  {\path{doi:10.1063/1.1498001}}.

\bibitem{Takayanagi:2017knl}
Tadashi Takayanagi and Koji Umemoto.
\newblock {Entanglement of purification through holographic duality}.
\newblock {\em Nature Phys.}, 14(6):573--577, 2018.
\newblock \href {http://arxiv.org/abs/1708.09393} {\path{arXiv:1708.09393}},
  \href {http://dx.doi.org/10.1038/s41567-018-0075-2}
  {\path{doi:10.1038/s41567-018-0075-2}}.

\bibitem{Jiang:2024ijx}
Xin Jiang, Peng Wang, Houwen Wu, and Haitang Yang.
\newblock {Alternative to purification in conformal field theory}.
\newblock {\em Phys. Rev. D}, 111(2):L021902, 2025.
\newblock \href {http://arxiv.org/abs/2406.09033} {\path{arXiv:2406.09033}},
  \href {http://dx.doi.org/10.1103/PhysRevD.111.L021902}
  {\path{doi:10.1103/PhysRevD.111.L021902}}.

\bibitem{Jiang:2025tqu}
Xin Jiang, Houwen Wu, Peng Wang, and Haitang Yang.
\newblock {Mixed State Entanglement Entropy in CFT}.
\newblock 1 2025.
\newblock \href {http://arxiv.org/abs/2501.08198} {\path{arXiv:2501.08198}}.

\bibitem{Jiang:2025dir}
Xin Jiang, Haitang Yang, and Zilin Zhao.
\newblock {Entanglement entropy of mixed state in thermal CFT2}.
\newblock {\em Phys. Rev. D}, 112(4):046025, 2025.
\newblock \href {http://arxiv.org/abs/2501.11302} {\path{arXiv:2501.11302}},
  \href {http://dx.doi.org/10.1103/bpzx-kdgq} {\path{doi:10.1103/bpzx-kdgq}}.

\bibitem{Wang:2017bym}
Peng Wang, Houwen Wu, and Haitang Yang.
\newblock {AdS$_{3}$ metric from UV/IR entanglement entropies of CFT$_{2}$}.
\newblock 10 2017.
\newblock \href {http://arxiv.org/abs/1710.08448} {\path{arXiv:1710.08448}}.

\bibitem{Wang:2018vbw}
Peng Wang, Houwen Wu, and Haitang Yang.
\newblock {Fixing three dimensional geometries from entanglement entropies of
  CFT$_{2}$}.
\newblock {\em Chin. Phys. C}, 49(2):025106, 2025.
\newblock \href {http://arxiv.org/abs/1809.01355} {\path{arXiv:1809.01355}},
  \href {http://dx.doi.org/10.1088/1674-1137/ad93b8}
  {\path{doi:10.1088/1674-1137/ad93b8}}.

\bibitem{Wang:2018jva}
Peng Wang, Houwen Wu, and Haitang Yang.
\newblock {Fix the dual geometries of $T\bar{T}$ deformed CFT$_{2}$ and highly
  excited states of CFT$_{2}$}.
\newblock {\em Eur. Phys. J. C}, 80(12):1117, 2020.
\newblock \href {http://arxiv.org/abs/1811.07758} {\path{arXiv:1811.07758}},
  \href {http://dx.doi.org/10.1140/epjc/s10052-020-08680-7}
  {\path{doi:10.1140/epjc/s10052-020-08680-7}}.

\bibitem{Jiang:2024xqz}
Xin Jiang, Peng Wang, Houwen Wu, and Haitang Yang.
\newblock {Realization of ''ER=EPR''}.
\newblock 11 2024.
\newblock \href {http://arxiv.org/abs/2411.18485} {\path{arXiv:2411.18485}}.

\bibitem{Synge:1960ueh}
J.~L. Synge, editor.
\newblock {\em {Relativity: The General theory}}.
\newblock 1960.

\bibitem{Buser2010}
Peter Buser.
\newblock {\em Geometry and spectra of compact Riemann surfaces}.
\newblock Birkh{\"a}user Boston, 2010.
\newblock \href {http://dx.doi.org/10.1007/978-0-8176-4992-0}
  {\path{doi:10.1007/978-0-8176-4992-0}}.

\bibitem{Brown1986}
J.~David Brown and M.~Henneaux.
\newblock Central charges in the canonical realization of asymptotic
  symmetries: An example from three-dimensional gravity.
\newblock {\em Commun. Math. Phys.}, 104:207--226, 1986.
\newblock \href {http://dx.doi.org/10.1007/BF01211590}
  {\path{doi:10.1007/BF01211590}}.

\bibitem{Tamaoka:2018ned}
Kotaro Tamaoka.
\newblock {Entanglement Wedge Cross Section from the Dual Density Matrix}.
\newblock {\em Phys. Rev. Lett.}, 122(14):141601, 2019.
\newblock \href {http://arxiv.org/abs/1809.09109} {\path{arXiv:1809.09109}},
  \href {http://dx.doi.org/10.1103/PhysRevLett.122.141601}
  {\path{doi:10.1103/PhysRevLett.122.141601}}.

\bibitem{Dutta:2019gen}
Souvik Dutta and Thomas Faulkner.
\newblock {A canonical purification for the entanglement wedge cross-section}.
\newblock {\em JHEP}, 03:178, 2021.
\newblock \href {http://arxiv.org/abs/1905.00577} {\path{arXiv:1905.00577}},
  \href {http://dx.doi.org/10.1007/JHEP03(2021)178}
  {\path{doi:10.1007/JHEP03(2021)178}}.

\bibitem{Wen:2021qgx}
Qiang Wen.
\newblock {Balanced Partial Entanglement and the Entanglement Wedge Cross
  Section}.
\newblock {\em JHEP}, 04:301, 2021.
\newblock \href {http://arxiv.org/abs/2103.00415} {\path{arXiv:2103.00415}},
  \href {http://dx.doi.org/10.1007/JHEP04(2021)301}
  {\path{doi:10.1007/JHEP04(2021)301}}.

\bibitem{Mori:2024gwe}
Takato Mori and Beni Yoshida.
\newblock {Does connected wedge imply distillable entanglement?}
\newblock 11 2024.
\newblock \href {http://arxiv.org/abs/2411.03426} {\path{arXiv:2411.03426}}.

\bibitem{Osborn:1991gm}
H.~Osborn.
\newblock {Weyl consistency conditions and a local renormalization group
  equation for general renormalizable field theories}.
\newblock {\em Nucl. Phys. B}, 363:486--526, 1991.
\newblock \href {http://dx.doi.org/10.1016/0550-3213(91)80030-P}
  {\path{doi:10.1016/0550-3213(91)80030-P}}.

\bibitem{Calabrese:2004eu}
Pasquale Calabrese and John~L. Cardy.
\newblock {Entanglement entropy and quantum field theory}.
\newblock {\em J. Stat. Mech.}, 0406:P06002, 2004.
\newblock \href {http://arxiv.org/abs/hep-th/0405152}
  {\path{arXiv:hep-th/0405152}}, \href
  {http://dx.doi.org/10.1088/1742-5468/2004/06/P06002}
  {\path{doi:10.1088/1742-5468/2004/06/P06002}}.

\bibitem{Jiang:2023loq}
Xin Jiang, Peng Wang, Houwen Wu, and Haitang Yang.
\newblock {Timelike entanglement entropy in dS$_{3}$/CFT$_{2}$}.
\newblock {\em JHEP}, 08:216, 2023.
\newblock \href {http://arxiv.org/abs/2304.10376} {\path{arXiv:2304.10376}},
  \href {http://dx.doi.org/10.1007/JHEP08(2023)216}
  {\path{doi:10.1007/JHEP08(2023)216}}.

\bibitem{Jiang:2025jnk}
Xin Jiang and Haitang Yang.
\newblock {Entanglement Entropy of Conformal Field Theory in All Dimensions}.
\newblock 6 2025.
\newblock \href {http://arxiv.org/abs/2506.02786} {\path{arXiv:2506.02786}}.

\end{thebibliography}
\onecolumngrid
\newpage
\begin{center}
	{\large{\bf Appendix}}
\end{center}

In this Appendix, we will show that  the $S_{\text{vN}}(A:B)$ in equation (\ref{eq:Renyi}) 
calculated by subtraction approach in CFT$_{2}$  coincides with the entanglement of purification (EoP) $E_{p}(A:B)$.

First note that our purpose is to compute  the entanglement between disjoint segments $A$ and $B$. 
In the procedure of EoP, one first obtains a mixed state $\rho_{AB}$ after tracing out $C$ and $D$. 
Then auxiliary systems $\bar{A}$ and $\bar{B}$ are introduced to purify  $\rho_{AB}$. 
This is totally equivalent to  replacing $C$ and $D$ with auxiliary systems from the very outset,
in which no  mixed state appears as an intermediate step and we always work on pure states.

\begin{figure}[h]
\includegraphics[width=0.45\textwidth]{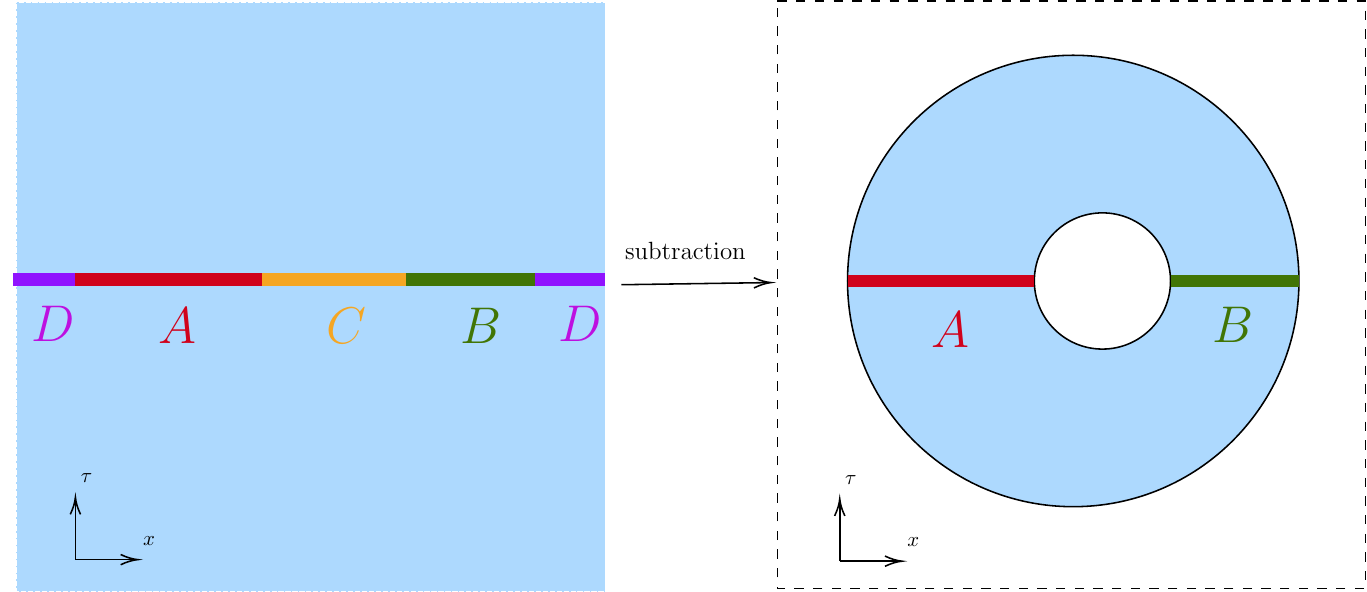}
\caption{After subtracting segments $C$ and $D$
with two discs in the infinite system, we obtain an asymmetric annular region in
which $A$ and $B$ are in a pure entangled state $\psi_{AB}$, which is identical to the annular CFT$_2$.
\label{fig:subtraction}}
\end{figure}

In order to compute the  entanglement between disjoint segments $A$ and $B$,
it is necessary to eliminate the overcounted degrees of freedom (denoted $D_{\text{oc}}$), which originate from the entanglement between the combined subsystem $AB$ and its complement $CD$. For instance, the EoP replaces $CD$ with auxiliary systems $\bar{A}\bar{B}$; its built-in minimization procedure guarantees the cancellation of $D_{\text{oc}}$. In any case, the purpose is to remove the effects caused by  $CD$ 
with a controlled manner. 
To this end, why not seek  a covariant approach to remove  $CD$  within the path integral, thereby directly eliminating $D_{\text{oc}}$? This is precisely what the SUBTRACTION procedure does. As illustrated in Figure \ref{fig:subtraction},
after removing segments $C$ and $D$ with two discs in the infinite system, we get the annular CFT in Figure \ref{fig:density}.

Notably, a striking advantage of taking the annular CFT$_2$  as the starting point  is that we can extend the 
method to higher dimensions straightforwardly. Specifically, in a recent paper \cite{Jiang:2025jnk}, by replacing $A$ and $B$ with $(D-1)$-dimensional disks,
we consider the solid torus $\mathbb{B}^{D-1}\times S^1$. 
We are then able to provide a very simple field-theoretic method to calculate the entanglement entropies of CFT in all dimensions.

We now provide specific arguments about the equivalence between $S_{\text{vN}}(A:B)$ and EoP.

\begin{figure}[h]
	\centering
	\includegraphics[width=0.65\textwidth]{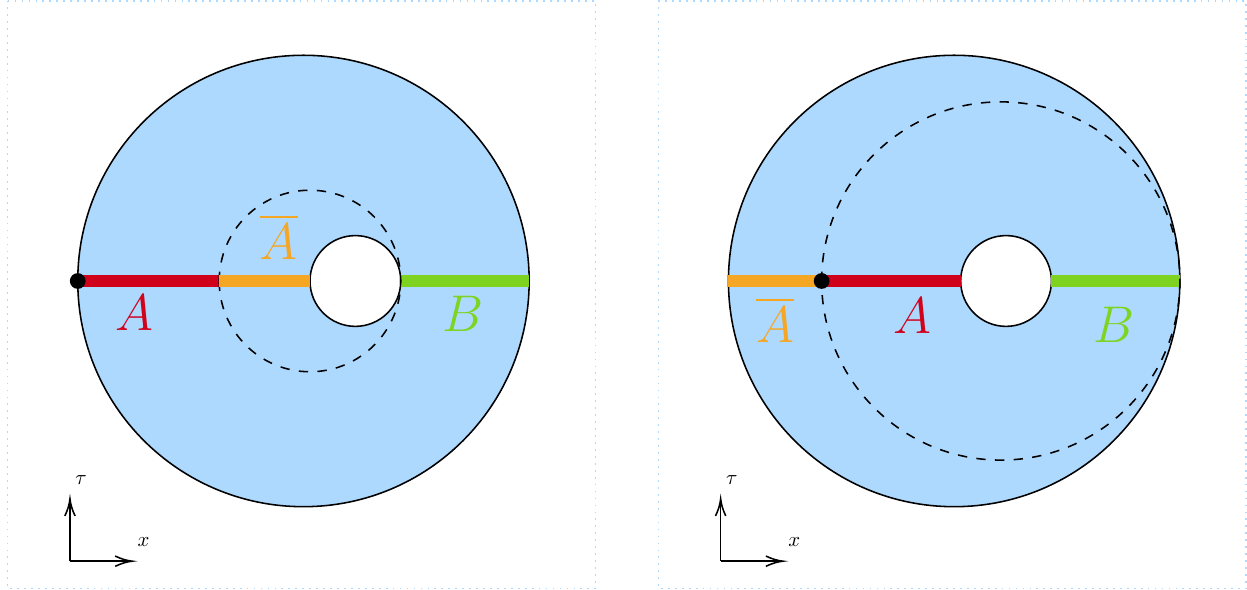}
	\caption{Two configurations for adding the auxiliary system $\bar{A}$. Each black point denotes the origin.\label{fig:EoP}}
\end{figure}

The entanglement of purification $E_{p}(A:B)$ is defined as
\begin{equation}
	E_{p}(A:B) = \min_{\bar{A},\bar{B}} S_{\text{vN}}(A\bar{A}:B\bar{B}) ,
\end{equation}
where $\bar{A},\bar{B}$ are auxiliary systems, and the minimization is over all possible purifications $\vert\Psi_{A\bar{A}B\bar{B}}\rangle$ of $\rho_{AB}$. 

Consider a static CFT$_{2}$ at $t=0$ with subsystems $A = (0,l_{A})$ and $B = (l_{A}+d,l_{A}+l_{B}+d)$. 
First, add an auxiliary system $\bar{A} = (l_{A}, l_{A} + l_{\bar{A}})$ (left panel of Fig.~\ref{fig:EoP}). 
From the results in \citep{Jiang:2024ijx}, it is straightforward to get
\begin{equation}
	\frac{\partial S_{\text{vN}}(A\bar{A}:B)}{\partial l_{\bar{A}}}=\frac{c}{6(d-l_{\bar{A}})}\sqrt{\frac{l_{B}(d+l_{A})}{(l_{\bar{A}}+l_{A})(d-l_{\bar{A}}+l_{B})}}\ge0,
\end{equation}
with $0<l_{\bar{A}}<d.$ One can also choose another auxiliary system
$\bar{A}=(-l_{\bar{A}},0)$, as shown in the right panel of Figure
\ref{fig:EoP}, where we still have
\[
\frac{\partial S_{\text{vN}}(A\bar{A}:B)}{\partial l_{\bar{A}}}=\frac{c}{6(d+l_{\bar{A}}+l_{A}+l_{B})}\sqrt{\frac{l_{B}(d+l_{B})}{(l_{\bar{A}}+l_{A})(d+l_{\bar{A}}+l_{A})}}\ge0,
\]
with $l_{\bar{A}}>0.$

Treating $A\bar{A}$ as a composite subsystem and adding $\bar{B}$ adjacent to $B$, with the same pattern, it follows that
\begin{equation}
	\frac{\partial S_{\text{vN}}(A\bar{A}:B\bar{B})}{\partial l_{\bar{B}}} \geq 0 \quad \text{and} \quad \frac{\partial S_{\text{vN}}(A\bar{A}:B\bar{B})}{\partial l_{\bar{A}}} \geq 0 .
\end{equation}
These inequalities imply that $S_{\text{vN}}(A\bar{A}:B\bar{B})$ is minimized when $l_{\bar{A}} \to 0$ and $l_{\bar{B}} \to 0$. 
Consequently, the entanglement of purification reduces to  
\[
E_{p}(A:B) = S_{\text{vN}}(A:B) .
\]


\end{document}